\documentclass[12pt]{article}%
 \usepackage{a4wide}
\usepackage[pdftex]{graphicx}
\usepackage[section]{placeins}
\usepackage{times}
\usepackage{amssymb,amsbsy,amsmath}
 \usepackage{cite}
\def\physrep{Phys. Rep.\ }
\def\pra{Phys. Rev. A\ }

\def\nphysa{Nucl. Phys. A\ }

\def\d{\mathrm{d}}

\begin{document}
\title{Exotic atoms in two dimensions\footnote{PACS: {36.10.Dr,03.65.Ge,03.65.nk}}}
\author{M Combescure$^1$\footnote{m.combescure@ipnl.in2p3.fr}, 
C Fayard$^1$\footnote{c.fayard@ipnl.in2p3.fr},
 A Khare$^2$\footnote{khare@iiserpune.ac.in},
 J-M Richard$^1$\footnote{j-m.richard@ipnl.in2p3.fr}\\[3pt]
{\small $^1${}Institut de Physique Nucl\'eaire de Lyon, Universit\'e de
Lyon, CNRS-IN2P3-UCBL,}\\[-2pt]
{\small 4,~rue Enrico Fermi, Villeurbanne, France}\\[3pt]
{\small $^2${}Raja Ramanna Fellow, IISER, Pune 411021, India}}
\date{\today}
\maketitle
\begin{abstract}
We study the behavior of energy levels in two dimensions for exotic atoms,
i.e.,  when a long-range
attractive potential is supplemented by a short-range interaction, and
compare
the results with these of the one- and three-dimensional cases. The
energy shifts are well reproduced by a scattering length formula
$ \delta{E}= A_0^2/\ln (a/R)$, where $a$ is the scattering length in the
short-range potential, $A_0^2/(2\,\pi)$ the square of the wave function at
the origin
in the external potential, and  $R$ is related to the derivative with
respect to the energy of the solution  that is regular at large distances.
\end{abstract}
\section{Introduction}
\label{sec:intro}
Hadronic atoms give valuable information about strong interactions at
low energy. For a review, see, e.g.,\cite{1996NuPhA.606..283G}. They have
also motivated several studies on the
behavior of the energy levels in a Schr\"odinger operator, with a potential
$V_1+\lambda\,V_2$, where $V_1$ dominates at large distances, but is
superseded
by $V_2$ at short distances. The case of exotic atoms corresponds to a world
with three dimensions, where $V_1=-1/r$ (as a negatively-charged hadron
orbits
near the nucleus and is almost unscreened by the remaining electrons, if
any),
and $V_2$ describes the short-range hadronic interaction. But the situation
is
far more general, and many features do not depend on the Coulomb character
of
$V_1$. Nevertheless, we shall use the word ``exotic atom'' for such a
system,
``atomic'' for the energy domain of the eigenstates of $V_1$ alone, and
``nuclear'' for any typical energy within $V_2$ alone, for the sake of
simplicity.

The spectral problem of exotic atoms
\cite{1996NuPhA.606..283G,1982PhR....82...31B,2007IJMPB..21.3765C}
differs significantly from the ordinary perturbation
theory, for which an expansion of the eigenenergies in powers of $\lambda$
is
attempted. For exotic atoms, the energies for $V_1+\lambda \,V_2$ are often
very
close to the ones for $V_1$ alone, but perturbation theory usually does not
hold. For instance, if $\lambda\,V_2$ is an infinite hard core of small
radius, the energies are
slightly shifted upwards, but the ordinary perturbative expansion diverges
already at the first order. The proper concept here is ``radius perturbation
theory'', as described by Mandelszweig \cite{1977NuPhA.292..333M}.

In this paper, we discuss how exotic atoms behave in $d=2$ dimensions. It
may be noted  that the study of exotic atoms in
$d=1$ is more straightforward, and already discussed in the literature
\cite{2007IJMPB..21.3765C}. The $d=2$ case is more delicate. The leading
order
term for the energy shift is easily identified, and linked to $\ln a$, where
$a$
is the scattering length in the short-range potential. As in the $d=3$ case,
the
overall coefficient is the square of the  wave-function at the origin in the
external potential. However, the scale regularizing this leading term, i.e.,
the
radius $R$ leading to $\ln a\to \ln(a/R)$ is not immediate, but it can be
derived from a matching of the solution of $V_1$ which  is
normalizable to the asymptotic solution emerging from the short-range term
$\lambda\,V_2$.

The case of $d=2$ dimensions is rather special in spectral problems, as it
corresponds to the largest value of $d$ for which an attractive potential,
however
weak, always holds at least one bound state,
see, e.g., \cite{yang:85,Grosse:847188}.
\footnote{More precisely, what is sufficient is that the integral of the
potential over the whole space is positive.}\@
Hence, for $d\le 2$, if $V_2$
is attractive, $\lambda V_2$ immediately develops its own bound state, which
becomes the ground state of the Hamiltonian. However, this process is less
effective for $d=2$ than for $d=1$, and the spectrum, as a function of
$\lambda$ evolves more slowly.

This paper is organized as follows. In Sec.~\ref{sec:3D} and
Sec.~\ref{sec:1D},
we give a brief reminder about the cases of $d=3$ and $d=1$ space dimensions
respectively, with particular emphasis on
the phenomenon of level rearrangement and on the scattering length 
(hereafter referred to as SL) formula for
the energy shifts. In Sec.~\ref{sec:2D}, we present the results for the
case of $d=2$ dimensions. The theoretical framework is presented in
Sec.~\ref{sec:deri}, before the final discussion in Sec.~\ref{sec:disc}.
\section{Exotic atoms in three dimensions}\label{sec:3D}
\begin{figure}[!htbc]
\centerline{\includegraphics[width=.30\textwidth]{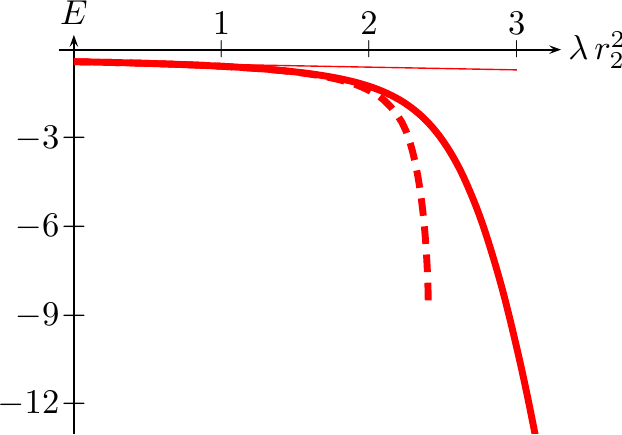}
\hspace*{.05\textwidth}
\includegraphics[width=.52\textwidth]{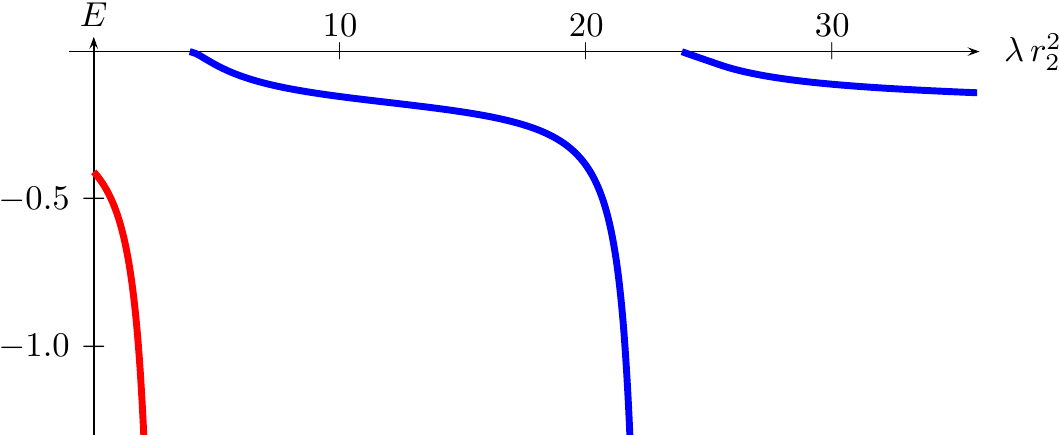}}
\caption{\label{fig:rea3D} Level rearrangement in three dimensions. A wide
and
weak external square well $V=-\lambda_1\, \Theta(r_1-r)$, with $\lambda_1=4$
and $r_1=1$, is supplemented by a short-range
square well of increasing depth, $\lambda\,V_2= -\lambda\,\Theta(r-r_2)$,
with $r_2=0.1$, in units
where
$\hbar^2/(2\mu)=1$, $\mu$ being the reduced mass. The first few energy
levels
are shown against $\lambda\,r_2^2$. Bottom: first three levels.
Top: magnification for ground-state alone, with the exact value (thick line)
compared to the first-order perturbation theory and SL formula. }
\end{figure}

There is an abundant literature about exotic atoms in three dimensions, 
motivated by experiments with pionic, kaonic and antiprotonic atoms
\cite{1996NuPhA.606..283G,1982PhR....82...31B}.
 The simplest model consists of a
two-component potential
\begin{equation}
\label{eq:2pot}V_1+\lambda \,V_2~,
\end{equation}
where $V_1$ is a long-range interaction with one or several bound states.
Genuine exotic atoms correspond to $V_1(r)\propto -1/r$. The second term,
with an explicit strength $\lambda$ introduced for the ease of the
discussion, accounts for the short-range interaction. The main results are:
\begin{itemize}
\item the shift is usually rather small, although $\lambda\,V_2$ can be very
large
at short distances,
\item the shift is usually well described by the approximate formula
\begin{equation}
\label{eq:DT} \delta{E}=E(\lambda)-E(0)\simeq 4\pi\,|\phi(0)|^2 a\,,
\end{equation}
where $\phi$ is the normalized wave function for $\lambda=0$, and $a$ the
scattering length in $V_2$ alone. In case $V_1$ is Coulombic, one
recovers the well-known SL formula by Deser, Golberger, Bauman and Thirring,
and Trueman
\cite{1954PhRv...96..774D,1961NucPh..26...57T}
\begin{equation}\label{eq:DT-Coul}
 \frac{E_n-E_n^{(0)}}{E_n^{(0)}}\simeq -\frac{4 a}{n B}\,,
\end{equation}
where $B$ is the Bohr radius and $n$ the principal number for the energy
$E_n^{(0)}$ in $V_1$ alone or $E_n$ in the total potential.
Many improvements and further corrections to this formula
have been discussed in the literature
\cite{1992ZPhyA.343..325C,2001JPhG...27.1421M}.
\item When $\lambda$ is varied, the shift usually varies very slowly, except
near the specific values $\lambda_1$, $\lambda_2$, \dots, where the energy
levels
change very rapidly, and a \emph{level rearrangement} occurs: near
$\lambda=\lambda_n$, the $n^{\rm th}$ energy drops toward very large
(negative)
values in the nuclear domain, and is replaced in the upper part of the
spectrum by the next level,
which in turn is replaced by the next one, etc. An example is given is
Fig.~\ref{fig:rea3D}. Further examples are provided, e.g., in
\cite{2007IJMPB..21.3765C}. The critical values $\lambda_n$ correspond to
the
coupling thresholds for which the short-range interaction $\lambda\,V_2$
starts
supporting a first or an  additional bound state.
\end{itemize}

\section{Results in one dimension}\label{sec:1D}
\begin{figure}[!htbc]
\begin{center}
\includegraphics[width=.28\textwidth]{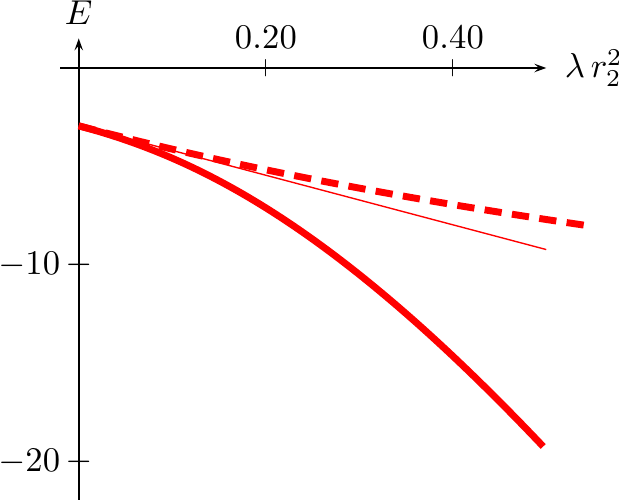}\hspace*{.1\textwidth}
\includegraphics[width=.52\textwidth]{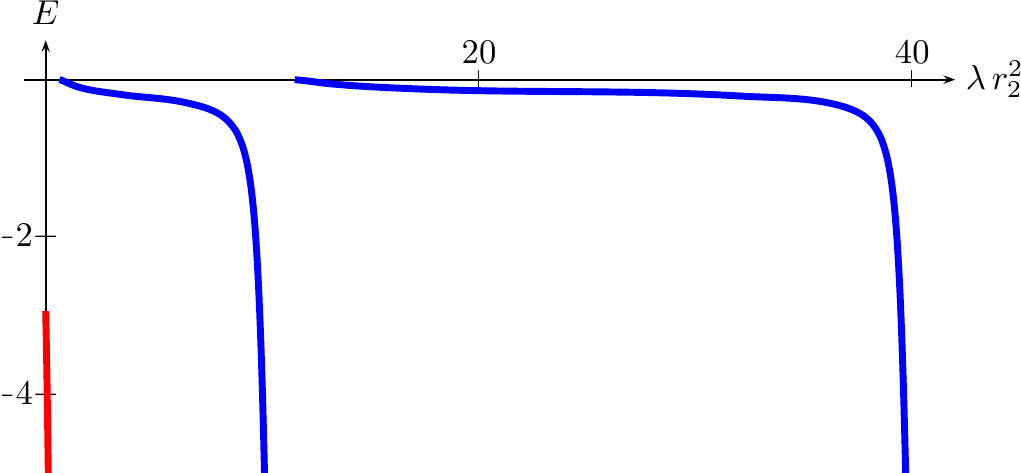}
\end{center}
 \caption{\label{fig:rea1D}Level rearrangement in one dimension. Same
interaction and same notation as Fig.~\ref{fig:rea3D}.
}
 \end{figure}

An example of spectrum of exotic atom in $d=1$ is shown in
Fig.~\ref{fig:rea1D}. It
consists again of a superposition of two square wells, the strength of the
short-range one being varied. The main
differences, as compared to the more familiar $d=3$ case are:
\begin{itemize}
 \item
As soon as $\lambda$  slightly departs from zero, the atomic ground state
energy immediately drops towards the range of  the nuclear energies.
\item
As a coupling threshold in $\lambda\,V_2$ is reached and $\lambda$ further
increases, a plateau
is observed; the corresponding energy drops, and, by rearrangement, a upper
level makes another plateau near the same value. This plateau in the
sector of the \emph{even} parity states, corresponds to an unperturbed
energy level in the \emph{odd} sector of $V_1$. Indeed, the orthogonality
with the ground state
forces  a zero in the wave function near $x=0$, and mimics an odd state.
\item
The Deser--Trueman formula, if translated for $d=1$,  reads
\begin{equation}
 \delta E \simeq -2\,\frac{|\phi(0)|^2}{a}~.
\end{equation}
The presence of the scattering length $a$ in the denominator can be
understood
by dimensional analysis. Also, weaker the short-range interaction
$\lambda\,V_2$, more flat the zero-energy wave function, and thus
larger the scattering length $a$, defined (as for $d=3$) as the
abscissa where the asymptotic zero-energy wave function vanishes.
\end{itemize}

\section{Results for two dimensions}\label{sec:2D}
\begin{figure}[!htbc]
\begin{center}\includegraphics[width=.4\textwidth]{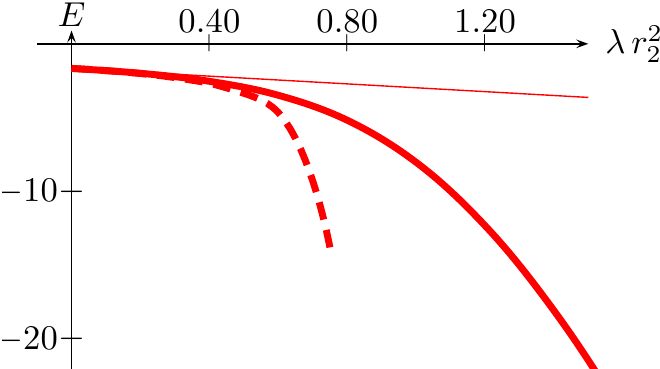}
\hspace*{.1\textwidth}
\includegraphics[width=.4\textwidth]{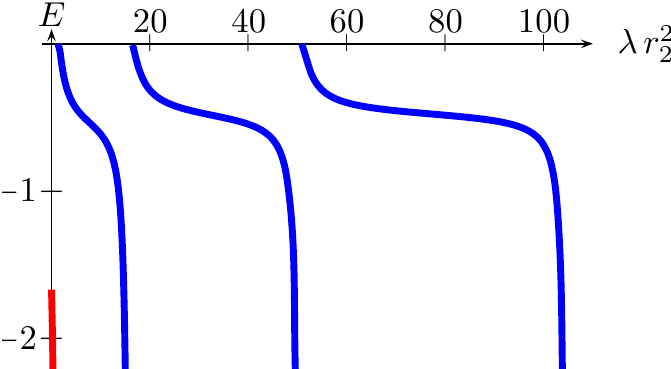}
\end{center}
\caption{\label{fig:rea2D} Level rearrangement in two dimensions, for the
first four levels. Same interaction and same notation as
Fig.~\ref{fig:rea3D}.}
\end{figure}

The calculation can be repeated for the isotropic (i.e., azimuthal quantum
number $m=0$) states
with $d=2$. If the atomic spectrum is examined for increasing values of the
strength of the
short-range interaction, a pattern of level rearrangement is clearly
identified, see Fig.~\ref{fig:rea2D}.

\begin{figure}[!htbc]
\centering\includegraphics[width=.33\textwidth]{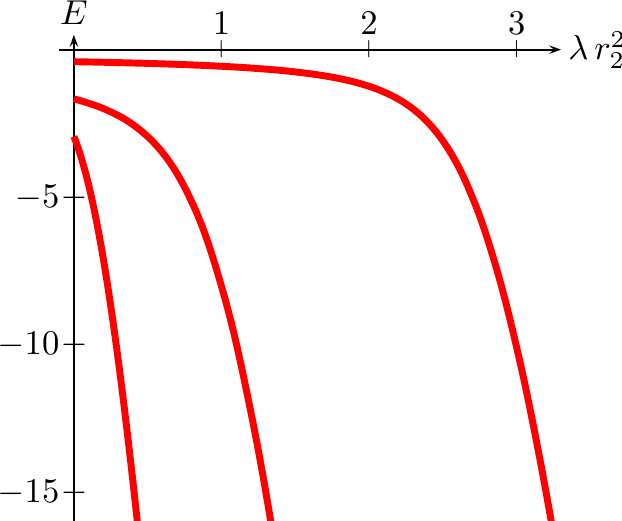}
\caption{\label{fig:rea2D1} Ground-state energy for $d=1$ (left), $d=2$ (middle)
 and $3$ (right) dimensions, with the same interaction as in the
previous figures.}
\end{figure}

The behavior of the ground state is
displayed again in Fig.~\ref{fig:rea2D1}, where it is compared to the $d=1$
and $d=3$ cases.
The trend is clearly intermediate between the plateau of $d=3$ and the
immediate fall-off of $d=1$.

\begin{figure}[!hbct]
\begin{center}
\includegraphics[width=.35\textwidth]{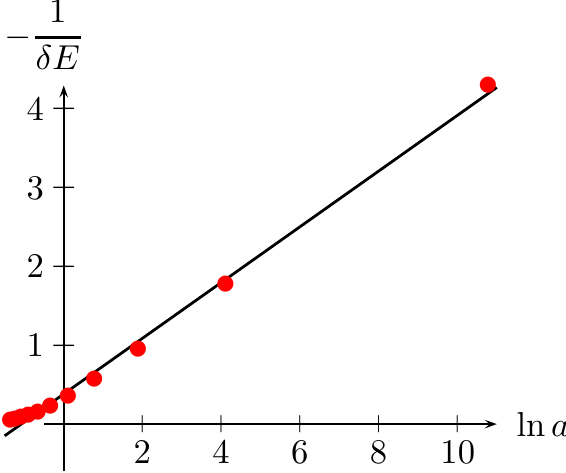}
 \end{center}
\caption{\label{fig:2Dfit} $1/\delta{E}$ against $\ln a$ for the
ground-state
energy in the double square-well of Fig.~\ref{fig:rea2D1}. The linear fit
cannot be distinguished from the exact results.}
\end{figure}

For small values of $\lambda$, we can easily identify the following behavior
for the $d=2$
energy shift $\delta E$
\begin{equation}
 \delta{E} = \frac{A_0^2}{\ln(a/R)}~.
\end{equation}
If one plots, as in the example shown in Fig.~\ref{fig:2Dfit}, $-1/\delta$
as a function of $\ln a$, one hardly distinguishes the exact values from the
results of a linear fit.

As discussed below, $A_0^2\simeq 2\pi\,|\phi(0)|^2$ and $a$ is the $d=2$
scattering length, as recently
revisited \cite{2008PhRvA..78e2711A,2009JMP....50g2105K}. The value of $R$
is
found of the order of magnitude of the ``Bohr radius'' of the wave function
in
the external potential, that is to say, the average radius. Its expression
is derived in the next section.

\section{Derivation of the energy shift}\label{sec:deri}
\subsection{General formula}
There are many approaches to the SL formula for $d=3$, and various
corrections and generalizations, see, e.g.,
\cite{1954PhRv...96..774D,1961NucPh..26...57T,1977NuPhA.292..333M,1992ZPhyA.343..325C,2001JPhG...27.1421M,2007IJMPB..21.3765C}
 and references there. For the $d=2$ case,
the following simple minded derivation is just based on the matching condition 
between
solutions of the Schr\"odinger equation  that are regular at short and at 
large distances.

For the sake of clarity, one can identify several approximations that are
made  when solving the bound-state problem in the potential
$V_1+\lambda\,V_2$:
\begin{enumerate}
 \item $V_1$ dominates at large distances
\item the energy $E$ in $V_1$ alone is a smooth function of the boundary
condition enforced at $r=0$,
\item $V_1$ and the energy term $E\,u$ can be neglected at very small
distances, where $\lambda\,V_2$ dominates.
\end{enumerate}

Let us start with Schr\"odinger equation for the external potential alone, 
i.e.,
\begin{equation}\label{eq:DTT1}
-u''-\frac{u}{4\,r^2}+ V_1\,u-E\,u=0~,
\end{equation}
where $u$ is the reduced radial wave function and we are working in the 
units with $\hbar=2m=1$. We denote $h(E,r)$ the
solution that is regular at infinity, i.e., $h(E,r)\propto
\sqrt{r}\,K_0(k\,r)$ at large distance, with $E=-k^2$ and $K_0$ the usual
Bessel function. The case of a confining interaction $V_1$ is treated later.
At short distance, this solution behaves as
\begin{equation}\label{eq:DTT2}
\begin{aligned}
 h(E,r)&=B(E)\,\sqrt{r}\,\ln r+A(E)\, \sqrt{r}+\cdots~,\\
h(E_0,r)&=A_0\,\sqrt{r}+\cdots
\end{aligned}
\end{equation}
for the modified energy $E$ and the unperturbed one $E_0$.
The unperturbed energy corresponds to $B(E_0)=0$, i.e., a solution that is
regular as $r\to 0$, and is normalized, leading to a real value $A_0=A(E_0)$
at energy $E_0$, that can be chosen to be positive.
For $E\neq E_0$ in the neighborhood  of $E_0$, we impose that the solution
remains normalized, i.e.,
\begin{equation}\label{eq:DTT3}
 \int_0^{+\infty} h(E,r)^2\,\d r=1~.
\end{equation}

By combining \eqref{eq:DTT1} for $h(E,r)$ and $h(E_0,r)$, one  obtains the
exact relation
\begin{equation}\label{eq:DTT4}
 A_0\,B(E)=(E-E_0) \int_0^{+\infty} h(E,r)\,h(E_0,r)\,\d r~
\end{equation}
which gives for the energy shift $\delta  E=E-E_0$ a first relation $\delta
E\simeq B\,A_0$.
It is rather precise.  Indeed, if the solution is kept to be normalized as
per \eqref{eq:DTT3}, and if $h(E,r)\to h(E_0,r)$ as $E\to E_0$, the integral
of $h(E,r)\,h(E_0,r)$ entering \eqref{eq:DTT4} is also equal to $1$, up to
second order in $\delta  E$.

Now,  $A=A_0+\tilde A_0\,\delta E+\cdots$,
one can identify the short-range behavior of $h(E,r)$ with $\sqrt{r}\,[\ln r
-\ln a]$, to obtain
\begin{equation}\label{eq:DTT5}
 \frac{B}{1}=\frac{A_0+\tilde A_0\,\delta E}{-\ln a}~,
\end{equation}
which when combined with  $B\simeq \delta E/A_0$ gives
\begin{equation}\label{eq:DTT6}
 \delta E\simeq- \frac{A_0^2}{\ln a+ A_0\,\tilde A_0}~.
\end{equation}
where the denominator can be cast as $\ln a- \ln R$.
This relation gives explicitly  the link between the energy and the boundary
condition at $r=0$,
expressed by $\ln a$, where $a$ is the scattering length in the short-range
interaction alone, in terms of the
quantities $A_0$ and $\tilde A_0$ linked to the value of the unperturbed
solution at the origin.

The effective range correction to the scattering length approximation can be
worked out explicitly, but turns out to be very small in most cases. For a
positive energy $E=k^2$, the $m=0$ solution to the scattering problem in
$\lambda\,V_2$ alone is
\cite{0305-4470-17-3-020,2008PhRvA..78e2711A,2009JMP....50g2105K}
\begin{equation}\label{eq:DTT7}
 u_2(k,r)=\sqrt{r}\left[\cot\delta(k)\,J_0(k\,r)+Y_0(k\,r)\right]~,
\end{equation}
where $J_0$ and $Y_0$ are Bessel functions, and
\begin{equation}\label{eq:DTT8}
\cot\delta(k)=\frac{2}{\pi}\left[\ln(a\,k/2)+\gamma\right]+\frac{1}{2}\,r_0\,
k^2+\cdots~
\end{equation}
involving the scattering length $a$ and effective range $r_0$. This
expression is easily translated for negative energies $E=-k^2$, and, anyhow,
the range of $E$ which is explored is small as compared to the typical
energy scale in the short-range potential, and thus the solution coming out
is safely approximated by its small $E$ limit
\begin{equation}\label{eq:DTT9}
 u_2=\sqrt{r}\,(\ln r + \ln a \mp \pi\,r_0\,E/4)~.
\end{equation}
This means one can replace $\ln a$ by $\ln a+\pi\,r_0 E_0/4$  to probe the
contribution of the effective range, which turns out negligible, provided
the energy shift $\delta E$ remains small as compared  to $E_0$.
\subsection{First example: double delta-shell}
To illustrate \eqref{eq:DTT6}, we consider as long range interaction an
attractive delta-shell of  strength $g_1$ and radius that can be set to
$R_1=1$ to fix the length scale. The solution can be worked out
analytically, in particular $\sqrt{r}\,K_0(k\,r)$ is regular at large $r$
and $\sqrt{r}\,I_0(k\,r)$ at small $r$. The delta-shell interaction imposes
the continuity of the radial solution $u$ near $r=R_1$ and the proper step
in its derivatives, to fix the unperturbed energy $E_0$. A second
delta-shell can be implemented at $r=R_2\ll R_1$, leading to an explicit
transcendental equation for the exact energy $E$, and shift $\delta
E=E-E_0$, to be compared to the simple approximate value $\delta
E'=-A_0^2/\ln a$ and the improved $\delta E''$ given by \eqref{eq:DTT6}. For
$g_1=1/2$, $R_2=0.04$, and $g_2=0.1$, one gets
\begin{equation}\label{eq:DDS1}
\begin{array}{ccc}
 \delta E &\delta E'&\delta E''\\
\ -0.000833975\ &\ -0.0008265\ &\ -0.0008340\
\end{array}
\end{equation}
i.e., an almost perfect agreement, when the $\ln R$ correction is taken into
account.
\subsection{Double exponential well}
\begin{figure}[htp]
\begin{center}
\includegraphics[width=.35\textwidth]{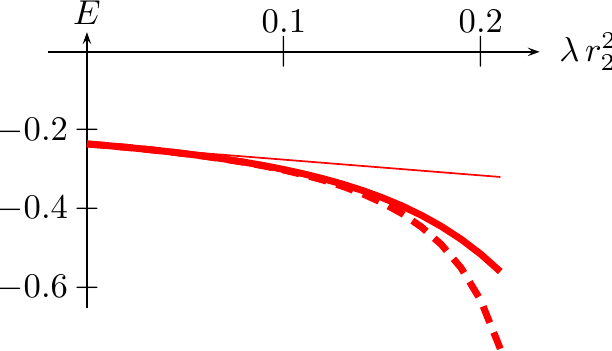}
\end{center}
 \caption{Exponential well supplemented by another exponential of shorter
range. Thick line: exact, dashed line: SL formula, thin line: perturbation
theory. We use here $V=-g_1 \, \exp(-r/r1) - \lambda\,\exp(-r/r_2)$ with
$r_1=2$, $g_1=1$ and $r_2=0.02$.}
 \label{fig:dbexp}
\end{figure}

As an example involving smooth potentials, we consider the exponential
potential 
$V_1=-g_1 \, \exp(-r/r1)$ with $g_1=1$ and take $r_1=2$ for the long-range
interaction, and study the changes due to another exponential interaction,
$-\lambda\,\exp(-r/r_2)$ with a much shorter range $r_2=0.02$ and a variable
strength. The results are displayed in Fig.~\ref{fig:dbexp}. Again, there is
a net gain as compared to the ordinary perturbation theory, and a good 
agreement  with the exact calculation as long as the deviation from the unperturbed energy is not too large.

\subsection{Harmonic confinement}
The problem is to study how the levels, in particular the ground state,  are
modified when a $d=2$ harmonic oscillator is supplemented by a short-range
interaction. A recent contribution is by Farrell and van Zyl
\cite{2010JPhA...43a5302F}. They first stressed the property of
\emph{universality}, namely that the energy shift does not depend on the
details of the short-range potential, but instead is governed by the
scattering length alone. This is, indeed, a very general property of the exotic
atoms, in the general sense define in the introduction
\cite{2007IJMPB..21.3765C}. For $V_1=r^2$, the general solution that is
regular at large distances can be written as
\begin{equation}\label{eq:HO1}
h(k,r)=\exp(-r^2/2)\, \sqrt{r}\, U(1/2-k^2/4,1,r^2)~,
\end{equation}
in terms of the confluent hypergeometric function $U$. This expression is
simpler, but equivalent to the one given in \cite{2010JPhA...43a5302F}. From
\eqref{eq:HO1}, one can calculate explicitly the normalization integral
$I(E)$ and its derivative.
The short-range behavior of $h(k,r)$ is known and if the ratio of the
$-\sqrt{r}$ to $\sqrt{r}\,\ln r$ coefficients is identified with $\ln a$, one
recovers the formula given in \cite{2010JPhA...43a5302F}. Our prescription
\eqref{eq:DTT6} corresponds to an approximate, but accurate, solution to the
matching equation. For instance, using $\lambda\,V_2=-\lambda\,\exp(-r/R_2)$
as an additional potential, with $R_2=0.02$ and $\lambda=-80$, one gets,
using the same notation as above and $\tilde \delta E$ for
\cite{2010JPhA...43a5302F},
\begin{equation}\label{eq:HO2}
\begin{array}{cccc}
  \delta E&\tilde\delta E&\delta E'&\delta E''\\
\ -0.06999\ &\  -0.07017\ &\ -0.07092\ &-0.07020
\end{array}
\end{equation}
Clearly, the main discrepancy comes from reducing this short-range
interaction to a zero-range ansatz. Once this is accepted,  our approximate
treatment is nearly exact as compared to the precise matching of $h(k,r)$ to
the $\sqrt{r}\, \ln (r/a)$ boundary condition.

\section{Summary}\label{sec:disc}
In this  note, we have studied how the energy levels in a wide potential
are modified by a
short-range attraction of increasing strength, focusing on the case of $d=2$
space dimensions, as compared to the $d=1$ and $d=3$ situations.

The energy shifts in a given external potential are well described by the
following SL formulas,
\begin{equation}\label{eq:fit}
 \delta{E}=\left\{
  \begin{aligned}
A_0^2/a &\quad (d=1)\\
A_0^2/\ln(a/R) &\quad (d=2)\\
A_0^2\, a &\quad (d=3)
\end{aligned}
\right.
\end{equation}
i.e., a perfect fit is obtained if $A_0^2$ (and $R$ for $d=2$) are treated
as free
parameters. Moreover,  $A_0$ can be identified with the first non vanishing
coefficient of the short-range expansion of the radial wave function and is
thus proportional  to
$ \phi(0)$,  the wave function at the origin for the
state in the external potential alone.
The ratio is $A_0^2/|\phi(0)|^2= 2 \ (d=1), 2\,\pi\ (d=2),\ 4\,\pi \ (d=3)$,
the unit-sphere area in $d$ dimensions.

In the $d=2$ case which is our main concern, a formula has been derived for
$R$, namely
$\ln R= - A_0\, A'(E_0)$, where $A(E)$ is the coefficient of $\sqrt{r}$ in
the normalized wave function, assumed to be real and positive and to match
$A_0$ at energy $E_0$.

This SL relation becomes more accurate when
additional potential $V_2$ becomes more short-ranged. In particular, it 
improves significantly the simple prediction from first order perturbation 
theory in $d=2$ and $d=3$.

This study of exotic atoms is intimately linked to the statistical physics
of
bosons. The common tool is the \emph{pseudo-potential}, which enables one to
replace a finite (but short) range interaction by a contact interaction.
Deriving the pseudo-potential as a function of the scattering length for
different values of the space dimension $d$ has been extensively discussed.
The case of $d=2$ is notoriously delicate, see, e.g.,
\cite{2008EL.....8440001Y,2002PhRvA..66a5601L,2002PhRvL..88a0402O,
2010JPhA...43a5302F} for recent contributions.
\subsection*{Acknowledgements}
This work was done with the support of the Indo--French cooperation
program CEFIPRA 3404-4.
%

\end{document}